      [1999/12/01 v1.4c Il Nuovo Cimento]
\documentclass{cimento}

\title{The Electro-Weak model as low-energy sector of 8-dimensional General Relativity}
\author{F.~Cianfrani\from{ins:x}\ETC,
G.~Montani\from{ins:x}\from{ins:y}}
\instlist{\inst{ins:x} ICRA-International Center for Relativistic Astrophysics\\
Dipartimento di Fisica (G9), Universit\`a  di Roma, ``La Sapienza", Piazzale Aldo Moro 5, 00185 Rome, Italy.
  \inst{ins:y} ENEA C.R. Frascati (Dipartimento F.T.P.), via Enrico Fermi 45, 00044 Frascati, Rome, Italy.}
\PACSes{{00.00}}
\begin{document}

\maketitle

\begin{abstract}
In a Kaluza-Klein background $V^4\otimes S^3$, we provide a way to reproduce, by the dimensional reduction, a 4-spinor with a $SU(2)$ gauge coupling. Since additional gauge violating terms cannot be avoided, we compute their order of magnitude by virtue of the application to the Electro-Weak model. \end{abstract}

\section{Layout}
The development of a unified theory for all interactions is one of the main issue of Theoretical physicists. In this sense, a great success has been obtain by the gauge approach; 
hence, the unification is performed simply by enlarging the gauge group, such that it contains those of Strong and Electro-Weak models.\\ 
However, this procedure does not work for gravity, since to recast it as a gauge theory is highly non-trivial. 
A different approach consists in promoting the internal space of gauge transformations to a part of the space-time itself, so it relays on the introduction of extra-dimensions. The standard way to this task is that of Kaluza-Klein (KK) theories\cite{2,3}.\\
KK models \cite{4,6} are based on considering a space-time manifold which is the direct sum of a 4-dimensional one $V^4$ and of an extra-dimensional manifold $B^K$, which has to be compact and homogeneous. The compactness is required in order to explain it undetectability, in fact one assumes it to be compactified to distances not yet available in current experiments (it implies its length scale to be $<10^{-17}cm$). Instead, the homogeneity allows to reproduce the algebra of a gauge group by that of the Killing vectors $\xi_{\bar{M}}$ of $B^K$, in the following way
\begin{equation}
\label{a1} \xi^{n}_{\bar{N}}\partial_n
\xi_{\bar{M}}^{m}-\xi^{n}_{\bar{M}}\partial_n
\xi_{\bar{N}}^{m}=C^{\bar{P}}_{\bar{N}\bar{M}}\xi^{m}_{\bar{P}}
\end{equation}
being $C^{\bar{P}}_{\bar{N}\bar{M}}$ structure constants. In what follows, we shall indicate with greek letters  $\mu,\nu$ and latin letters $m,n$ 4- and extra-dimensional coordinates, respectively.\\
Hence, we can write the metric tensor as 
\begin{equation}
\label{c1}
j_{AB}=\left(\begin{array}{c|c}g_{\mu\nu}+\alpha^2\gamma_{mn}\xi^{m}_{\bar{M}}\xi^{n}_{\bar{N}}A^{\bar{M}}_{\mu}A^{\bar{N}}_{\nu}
& \gamma_{mn}\xi^{m}_{\bar{M}}A^{\bar{M}}_{\mu} \\
\hline
\gamma_{mn}\xi^{n}_{\bar{N}}A^{\bar{N}}_{\nu}
& \alpha^2\gamma_{mn}\end{array}\right)
\end{equation}         
with $\alpha$ giving the length scale of $B^K$ (we assume it to be constant, but a more general treatment is possible).\\ 
Starting from the Einstein-Hilbert action for the above metric, i.e.
\begin{eqnarray*}
S=-\frac{c^{3}}{16\pi G_{(n)}}\int_{V^{4}\otimes B^{K}}
\sqrt{-j}R d^{4}xd^{K}y
\end{eqnarray*}
we obtain the following reduced action, in terms of $g_{\mu\nu}$ and $A^{\bar{M}}_{\mu}$, 
\begin{equation}
{}4\!^S=-\frac{c^{3}}{16\pi G}\int_{V^{4}}
\sqrt{-g}\bigg[R+R'_{N}+\frac{1}{4}\alpha^2 F^{\bar{M}}_{\mu\nu}F^{\bar{M}\mu\nu}\bigg]d^{4}x\label{lagspl}
\end{equation}
being $R$ and $R'$ scalars of curvature of $V^4$ and $B^K$, respectively, while for $F^{\bar{M}}_{\mu\nu}$ we have the expression
\begin{equation}
F^{\bar{M}}_{\mu\nu}=\partial_\nu A^{\bar{M}}_\mu-\partial_\mu A^{\bar{M}}_\nu+C^{\bar{M}}_{\bar{N}\bar{P}}A^{\bar{N}}_\mu A^{\bar{P}}_\nu.
\end{equation} 
The Lagrangian in (\ref{lagspl}) is the Einstein-Yang-Mills one, once we interpret $A^{\bar{M}}_{\mu}$ as gauge bosons; in this sense, we achieve the geometrization of a gauge bosonic component.\\
In the standard KK approach, spinors were introduced looking for the possibility to reproduce gauge transformations by extra-dimensional rotations\cite{bl2}. At first, under the KK hypothesis,  
eigenvalues of the Dirac equation on the extra-space behave like masses and they are of the compactification scale order; then, since no zero-eigenvalue state exists on a compact manifold \cite{13}, we always end up with 4-spinors, whose mass is of the Planck scale order (usual KK masses).\\
Moreover, the Atiyah-Hirzebruch theorem\cite{ah} rules out the possibility that the right-handed and left-handed lowest modes of the Dirac operator behave differently under n-bein rotations; this way, KK theories look unable to reproduce the chirality of the Electro-Weak (EW) model\footnote{These unpleasant issues can be solved by introducing external gauge bosons \cite{bl2}, but, this way, the most attractive feature of the KK approach (the geometrization of the bosonic component) is lost.}.\\ 
So we propose a different approach, where 4-spinors are the relic of the full field, after the dimensional reduction has been preformed\cite{6,7}. Therefore, we suggest that the dimensional reduction procedure determines in a non-trivial way properties of such fields. For example, due to the unobservability of the dynamics on $B^K$, 4-fields satisfy same equations as the 4+K-dimensional counterpart, but integrated on the extra-space\cite{7}. This procedure is justified by the idea that any observable must be averaged on the unobservable $B^K$. This new point of view allows to avoid the problem of large fermion masses.\\
Let us consider how this procedure works for the geometrization of $SU(2)$ gauge connections in a space-time $V^4\otimes S^3$\cite{06}. Since, we expect the compactification breaking up the full Lorentz group to the direct product of the 4-dimensional and the extra-dimensional ones, we develop a spinor representation as the product of ordinary spinors $\psi_1$, $\psi_2$ and a $SU(2)$ one, $\chi$, i.e. $\Psi=\sum_{s=1}^2\chi_{rs}\psi_s$. According with this assumption, we take non-Riemannian spinorial connections, such that in n-bein indices they read as follows
\begin{eqnarray}
\Gamma_{(\mu)}=^4{}\!\Gamma_{(\mu)}\qquad\Gamma_{(m)}=^3{}\!\Gamma_{(m)}=\frac{i}{2}\sigma_{(m)}
\end{eqnarray}
being $\sigma$ Pauli matricies. For $\chi$, we can take a form such that it is an approximate solution of the integrated Dirac equation, where additional terms are of $\beta^{-1}$ order, being $\beta$ a free parameter, i.e.
\begin{eqnarray}
\int_{S^{3}} d^{3}y\sqrt{\gamma}\gamma^{(m)}e_{(m)}^{m}\partial_{m}\chi=\frac{i}{2}\sigma_{(m)}\chi+O(\beta^{-1}).\label{avdir}
\end{eqnarray}
This form reads as follows  
\begin{equation}
\chi_{r}=\frac{1}{\sqrt{V}}e^{-\frac{i}{2}\sigma_{(p)rs}\lambda^{(p)}_{(q)}\Theta^{(q)}(y^{m})}\label{sp}
\end{equation}
with $V$ the volume of $S^3$ and the constant matrix $\lambda$ satisfies 
\begin{eqnarray}
(\lambda^{-1})^{(p)}_{(q)}=\frac{1}{V}\int_{S^{3}}
\sqrt{-\gamma}e^{m}_{(q)}\partial_{m}\Theta^{(p)}d^{3}y,
\end{eqnarray}
while for $\Theta^{(q)}$ we have
\begin{equation}
\label{theta}\Theta^{(p)}=\frac{1}{\beta}c^{(p)}e^{-\beta\eta}\qquad\eta>0
\end{equation}   
$\eta$ and $c^{(p)}$ being arbitrary functions on $S^3$. On this representation we impose the only request $\eta>0$, in order to deal with a small dependence of the spinor on extra-coordinates.\\
Because of the relation (\ref{avdir}), the form (\ref{sp}) allows to geometrize $SU(2)$ gauge connections, in fact, if we consider the dimensional reduction of the full Dirac equation
\begin{equation}   
S=-\frac{i\hbar c}{2}
\int_{V^{4}\otimes S^{3}}[\bar{\Psi}\gamma^{(A)}(\partial_{(A)}-\Gamma_{(A)})\Psi-c.c.]\sqrt{g}\sqrt{\gamma}d^{4}xd^{3}y=
\end{equation}
the right interaction between bosons and spinors arises, i.e. we get
\begin{eqnarray}
=-\frac{i\hbar c}{2}\int
\bigg[\bar{\psi}\gamma^{(\mu)}\bigg(D_{(\mu)}-\frac{i}{2\alpha}e^{\mu}_{(\mu)}W_{\mu}^{(m)}\sigma_{(m)}\bigg)
\psi-\frac{i}{2\beta}W_{\mu}^{(m)}M^{(n)}_{(m)}\bar{\psi}\sigma_{(n)}\psi-c.c.\bigg]\sqrt{-g}d^{4}x\nonumber
\end{eqnarray}
$M^{(n)}_{(m)}$ being
\begin{equation}
M^{(n)}_{(m)}\sigma_{(n)}=\frac{1}{V}\int_{S^{3}}d^{3}y \lambda^{(r)}_{(s)}e^{-\eta\beta}
e^{m}_{(m)}\partial_m c^{(s)}\int_0^1 ds\chi^{-(s-1)}\sigma_{(r)}\chi^{1-s}.
\end{equation}
The coupling constant $g$ can be introduced in this framework by a redefinition of gauge bosons $W^{(i)}_{\mu}\Rightarrow kg W^{(i)}_{\mu}$, and, in order the full Lagrangian density to coincide with the Yang-Mills one, the following relation must stand
\begin{equation}
\alpha^{2}=16\pi G(\hbar/gc)^{2}.\label{length}
\end{equation} 
We want to stress that the geometrization cannot be performed without the appearance of some gauge violating terms of $\beta^{-1}$ order. The request to regard them as negligible will give a lower bound for the parameter $\beta$ in the application of this scheme to the Standard Model.\\
Therefore, we now show that, in this framework, the geometrization of the Electro-Weak model ($SU(2)\otimes U(1)$) can be performed. The bosonic component simply requires to deal with $V^4\otimes S^3\otimes S^1$ space-time. For what concern spinors, the geometrization of a $U(1)$ gauge connection is easily obtained by a phase dependence on the $S^{1}$ variable $\theta$
\begin{equation}   
\Psi(x;\theta)=\frac{1}{\sqrt{\alpha'}}e^{in_{r}\theta}\psi(x)\qquad Y_{r}=\frac{n_{r}}{6}
\end{equation}
$Y_{r}$ being the hypercharge of the 4-dimensional spinor $\psi_{r}$ and $\alpha'$ the length of the extra-circle $S^1$.
However, the main difficulty to provide a consistent geometrization of the EW model relays on reproducing the chirality of the $SU(2)$ sector. Nevertheless we can overcome this obstacle, since in the present phenomenological approach we deal directly with 4­dimensional spinors.  In the case of a space-time manifold $V^4\otimes S^3$, for any quark generation and lepton family of the Standard Model, fermions can be obtained by the following multidimensional spinors 
 \begin{equation}   
\Psi_{Lr}(x;y;\theta)=e^{in_{s}\theta}\chi_{rs}(y)\psi_{Ls}(x)\qquad
\Psi_{Rr}(x;y;\theta)=e^{in_{r}\theta}\psi_{Rr}(x)\quad
\end{equation}
where $\psi_{R}$ and $\psi_{L}$ stand for the two four-dimensional chirality ($\gamma_{5}$) eigenstates.\\
But, since we develop our model in an 8-dimensional space-time, we deal with 16 components spinors. We suggest to recast them, so that each one contains a quark generation and a lepton family, i.e.
\begin{eqnarray*} 
\Psi_{L}=\frac{1}{\sqrt{V\alpha'}}\left(\begin{array}{c} \chi\left(\begin{array}{c} e^{in_{uL}\theta}u_{L}\\e^{in_{dL}\theta}d_{L}\end{array}\right)\\\chi\left(\begin{array}{c} e^{in_{\nu L}\theta}\nu_{eL}\\e^{in_{eL}\theta}e_{L}\end{array}\right)\end{array}\right)\qquad
\Psi_{R}=\frac{1}{\sqrt{V\alpha'}}\left(\begin{array}{c} \left(\begin{array}{c} e^{in_{uR}\theta}u_{R}\\e^{in_{dR}\theta}d_{R}\end{array}\right)\\\left(\begin{array}{c} e^{in_{\nu R}\theta}\nu_{eR}\\e^{in_{eR}\theta}e_{R}\end{array}\right)\end{array}\right). 
\end{eqnarray*}
This way, from 6 multidimensional spinors we are able to reproduce all Standard Model particles and an explanation for the equal number of quark generations and fermion families is provided.\\
Hence, after the dimensional reduction of the Dirac Lagrangian density, estimates obtained for $\alpha$ and $\alpha'$, from coupling constants $g$ and $g'$(\ref{length}), are as follows
\begin{equation}
\alpha=0.18\times 10^{-31}cm\qquad\alpha'=\alpha/\sin\theta_{w}=0.33\times10^{-31}cm.
\end{equation}
These estimates lead us to regard the stabilization of the extra-space as a quantum gravity process.\\
Additional $\beta^{-1}$ terms introduce unobserved interactions between the EW fields, which break down the gauge symmetries of the model.\\ 
From the observed partial width $\Gamma(n\rightarrow p+\nu_{e}+\bar{\nu}_{e})/\Gamma_{tot}$ associated to the decay of a neutron into a proton plus a couple of neutrino-antineutrino (violating the electric charge symmetry), we obtain $\beta\geq10^{14}$, taking into account the experimental limit \cite{10}
\begin{equation}
\Gamma(n\rightarrow p+\nu_{e}+\bar{\nu}_{e})/\Gamma_{tot}<8*10^{-27}.  
\end{equation}
Finally, the remaining feature of the Electro-weak model to be addressed in our scheme is the Higgs boson, which can be reproduced by the following multidimensional field
\begin{equation}
\label{Phi}\Phi=\left(\begin{array}{c}\Phi_{1}
\\\Phi_{2}
\end{array}\right)=\frac{1}{\sqrt{V\alpha'}}\chi\left(\begin{array}{c}e^{-3i\theta}\phi_{1}
\\ e^{3i\theta}\phi_{2}
\end{array}\right)
\end{equation}
subjected to the standard Higgs-like potential. After the dimensional reduction, the KK mass term is not suppressed and it can account for the fine-tuning we have to perform to stabilize Higgs mass. Moreover, Yukawa couplings between fermions and the Higgs field can reproduce a neutrino mass too. In fact, we have more freedom, with respect to the 4-dimensional EW model, in fixing the hypercharges of the Higgs components.\\  
Furthermore, even if massive gauge bosons are obtained, by the usual spontaneous symmetry breaking mechanism, nevertheless the following $\beta^{-1}$ corrections to the Lagrangian have to be taken into account, in the determination of masses
\begin{eqnarray*}
\ldots+2\beta^{-1}M^{(n)}_{(m)}W^{(m)}_{\mu}W^{(r)}_{\nu}g^{\mu\nu}\phi^\dag\sigma_{(n)}\sigma_{(r)}\phi+2\beta^{-1}N_{(m)(s)}W^{(m)}_{\mu}W^{(s)}_{\nu}g^{\mu\nu}\phi^{\dag}\phi+\ldots
\end{eqnarray*}
where $N_{(m)(s)}$ reads as follows
\begin{eqnarray*}
N_{(m)(s)}=\frac{1}{V\alpha'}\delta_{(q)(t)}\lambda^{(q)}_{(u)}\lambda^{(t)}_{(v)}\int d^{3}yd\theta \sqrt{\gamma}e^{m}_{(m)}\partial_m c^{(u)}e^{s}_{(s)}\partial_s\eta 
c^{(v)}e^{-2\beta\eta}\chi^{s}.
\end{eqnarray*}
From the current limit on the photon mass \cite{10}, $m_{\gamma}<6*10^{-17}eV$, we obtain a lower bound for $\beta$ greater than the previous one, i.e.
\begin{equation}
\frac{1}{\beta}\sim\frac{m_{\gamma}}{m_{Z}}\Rightarrow\beta>10^{28}.
\end{equation}


\begin{thebibliography}{99}

\bibitem{2}
\BY{Kaluza~T.}, 
\IN{Sitzungseber. Press. Akad. Wiss. Phys. Math. Klasse}{K1}{1921}{966}

\bibitem{3}
\BY{Klein~O.}\IN{Nature}{118}{1926}{516}

\bibitem{4}
\BY{Appelquist~T. \atque Chodos~A. \atque Frund~P.} \TITLE{Modern Kaluza-Klein
theories},  
          edited by \NAME{Addison Wesley Publishing Inc.}(1987)

\bibitem{6}
\BY{Cianfrani~F., Marrocco~A. \atque Montani~G.}
\IN{Int. J. Mod. Phys D}{14, 7}{2005}{1195}


\bibitem{bl2}
\BY{Bailin~D. \atque Love~A.}
\IN{\it Rep. Prog. Phys.}{50}{1987}{1087}

\bibitem{13}
\BY{Mecklenburg~W.}\IN{Phys. Rev. D}{26}{1982}{1327}

\bibitem{ah}
\NAME{Atiyah~M. \atque Hirzebruch~F.}
\TITLE{Essays on Topology and Related Topics}, 
                                           edited by Springer, Berlin (1970)

\bibitem{7}
\BY{Cianfrani~F. \atque Montani~G.}
\IN{\it Mod. Phys. Lett. A}{21, 3}{2006}{265}



\bibitem{06}
\BY{Cianfrani~F. \atque Montani~G.} \TITLE{Low-energy sector of 8-dimensional General Relativity: Electro-Weak model and neutrino mass},
submitted to Int. Journ. Mod. Phys. D

\bibitem{10}
\BY{Eidelman~S. et al. (Particle Data Group)} 
\IN{Phys. Lett.}{B592}{2004}{1}(URL: http://pdg.lbl.gov)


\end{thebibliography}
\end{document}